\documentclass[aps,twocolumn,showpacs]{revtex4}
\usepackage{dcolumn}
\usepackage{graphicx}
\usepackage{amsmath}
\usepackage{amsfonts}
\usepackage{amssymb}
\usepackage{psfrag}
\usepackage{wrapfig}
\usepackage{subfigure}
\usepackage{makeidx}
\usepackage{bm}
\usepackage{epsf}

\begin{document}

\title{ Temporal-Spatial Interference Pattern During Solitons Interaction}
\author{Li-Chen Zhao$^{1}$, Liming Ling$^{2}$, Zhan-Ying Yang$^{1}$, Jie Liu$^{3,4}$}\email{liu_jie@iapcm.ac.cn}
\address{$^1$Department of Physics, Northwest University, 710069, Xi'an, China}
\address{$^2$Department of Mathematics, South China University of Technology, Guangzhou 510640, China }
\address{$^3$Science and Technology Computation Physics Laboratory,
 Institute of Applied Physics and Computational Mathematics, Beijing 100088,
China}
\address{$^4$Center for Applied Physics and Technology, Peking University, 100084, Beijing,
China}
\date{August 12, 2013}
\begin{abstract}
Interference patterns associated with soliton-soliton interaction
are investigated in detail. We find that the temporal and spatial interference patterns exhibit
quite different characteristics. The period of the spatial
interference pattern is determined by the relative velocity of the
solitons, and the temporal pattern behavior is determined by the
peak amplitudes and the kinetic energy of the solitons.
Analytical expressions for the periods of the interference patterns
are obtained. A method for classifying the nonlinearity of many
nonlinear systems is proposed. As an example, we discuss
possibilities to observe these properties of solitons in a nonlinear
planar waveguide.
\end{abstract}
 \maketitle

\section{Introduction}

Since Zabusky and Kruskal introduced the concept of soliton at the
first time in 1965 \cite{Zabusky}, studies on dynamics of soliton have been
done in many fields, including hydrodynamics \cite{Barenblatt,Richardson}, quantum field
theory \cite{Barenblatt}, plasma physics \cite{Karpman}, nonlinear optics \cite{Yang,Serkin,Luo,Panomarenko,Z.Y. Yang}, and
Bose-Einstein condensate(BEC)\cite{Zhao,B. Wu,G.X. Huang,M. Matuszewski}. Nowadays, it is well known that
solitons collide elastically, and there are some phase shift
appearing after the collision \cite{Akhmediev,Voronvich,Genty,Onorato,Atre}, which demonstrate the
particle-like properties of soliton. These just provide us the knowledge about the properties of solitons before or after their collision. However, less attentions have been paid on the process of solitons' interaction. Considering soliton is
fundamentally a wave packet, one could expect that they can
interfere with each other during their interaction process \cite{Snyder,Kumar}. Notably, solitons' interaction
is a nonlinear process, since there is no linear superposition
principle for solitons \cite{Zabusky,Atre}. Then, are there any definite laws
in the nonlinear process? It deserves further research for soliton
application. Furthermore, it is known that one can measure many
physical quantities through light interference. What can we get
though the solitons' nonlinear interference pattern?

In this paper, we focus on the interference behavior in the process
of soliton-soliton interaction. We find that there are two types of
periodic behavior during the nonlinear interaction of solitons, namely,
separate temporal and spatial interference patterns. We derive the period expressions of interference pattern analytically and exactly. Based on the precise expressions of the two periods, one
can manipulate the interference pattern precisely through changing
the shape and velocities of the solitons. It is shown that a nonlinear
parameter could be obtained from interference pattern with
properties of solitons known.

The paper is organized as follows. In Section {\rm II}, we study on nonlinear interference pattern between solitons' interaction process in detail.
Definite scaling laws are found and the corresponding physics are
discussed. In Section {\rm III}, possibilities to observe
interference pattern are discussed in a planar waveguide. A method for measuring the nonlinearity of the nonlinear systems is proposed. A
conclusion is given in Section {\rm IV}.

\section{The two-soliton solution and their interference patterns}
Since nonlinear Schr\"odinger (NLS) equation can describe dynamics of soliton in many physical systems, such as nonlinear fiber \cite{Kibler}, Bose-Einstein condensate \cite{matterrw}, plasma system \cite{Bailung}, and even water wave tank \cite{Chabchoub},  we start with the
well-known simplified NLS equation
\begin{equation}
i\frac{\partial U(x,t)}{\partial t}+\frac{\partial^2
U(x,t)}{\partial x^2}+2g |U(x,t)|^2U(x,t)=0,
\end{equation}
which has been studied widely. The coefficient $g$ is the nonlinear
parameter. For a BEC system, it is related to scattering length, and
in nonlinear optics it is related to the nonlinear Kerr effect. The interaction between solitons can be studied based on multi-soliton solution which can be obtained by B\'{a}cklund transformation \cite{Mat,Cies,Dok}. Here, for  simplicity and without loss of generality, we will discuss two solitons' interaction based on the two-soliton solution. The
two-soliton solution can be presented as follows through nonlinear
superposition principle given by B\'{a}cklund transformation
\cite{Zhao,Mat}
\begin{equation}
    U=\frac{4F_1}{\sqrt{g}F_2},\quad
    |U|^2=\frac{[\ln(F_2)]_{xx}}{g},
\end{equation}
where
\begin{eqnarray*}
F_1&=&\left\{\,i a_{{1}}\left[ (b_1-b_2)^2+{a_{{1}}}^{2
}-{a_{{2}}}^{2} \right] \cosh(2X_2)\right. \nonumber\\
&&\left.+2\,a_{{1}}a_{{2}}
\left(b_{{2}}-b_{{1}} \right)\sinh(2X_2)\right\}e^{2{\rm
i}Y_1}\nonumber\\&&+\left\{\,i a_{{2}}\left[ (b_1-b_2)^2+{a_{{2}}}^
{2}-{a_{{1}}}^{2} \right] \cosh(2X_1)\right. \nonumber\\
&&\left.+2\,a_{{1}}a_{{2}}
(b_{{1}} -b_{{2}})\sinh(2X_1) \right\}e^{2{\rm i}Y_2},          
\end{eqnarray*}
\begin{eqnarray*}
F_2&=& \left[(a_1+a_2)^2+(b_1-b_2)^2\right] \cosh A_1\nonumber\\&&+
\left[(a_1-a_2)^2+(b_1-b_2)^2 \right] \cosh A_2 -4\,a_{{1}}a_{{2
}}\cos A_3,
\end{eqnarray*}
\begin{eqnarray*}
     X_{{1}}&=&a_{{1}} \left(x-4\,b_{{1}}t \right) +c_{{1}},\quad Y_{{1}}=b_{{1}}x+ 2\left( \,{a_{{1}}}^{2}-\,{b_{{1}}}^{2} \right) t+d_{{1}},\\
     X_{{2}}&=&a_{{2}} \left(x-4\,b_{{2}}t \right) +c_{{2}},\quad Y_{{2}}=b_{{2}}x+ 2\left( \,{a_{{2}}}^{2}-\,{b_{{2}}}^{2} \right) t+d _{{2}},\\
     A_1&=&2\left(\,a_{{1}}-\,a_{{2}} \right) x+ 8\left(\,a_{{2}}b_{{2}}-\,a_{{1}}b_{{1}} \right) t+2(\,c_{{1}}-\,c_{{2}}),\\
     A_2&=&2\left( \,a_{{2}}+\,a_{{1}} \right) x-8\left(\,a_{{1}}b_{{1}}+\,a_{{2}}b_{{2}} \right) t+2(\,c_{{2}}+\,c_{{1}}),\\
     A_3&=&2\left(\,b_{{1}}-\,b_{{2}} \right) x+4 \left(\,{a_{{1}}}^{2} -\,{a_{{2}}}^{2}+\,{b_{{2}}}^{2}-\,{b_{{1}}}^{2} \right)
     t\nonumber\\&&+2(\,d_{{1}}-\,d_{{2}}),
\end{eqnarray*}
$a_1$, $a_2$, $b_1$, $b_2$, $c_1$, $c_2$, $d_1$, $d_2$ are arbitrary
real numbers.
 When the related parameters are chosen, the solution will
present us the dynamics of two soliton directly. The collision of
them can be observed conveniently. As usual, one can observe that the
collision is elastic, and there are phase shifts appearing. These
particle-like characters have been studied in \cite{Akhmediev,Voronvich,Genty,Onorato,Atre}, which just provide us the knowledge about the properties of solitons before or after their collision.  Then, what about the interaction process of two solitons' collision? Interestingly, we find that
interference pattern can emerge during the process, which demonstrates the
wave properties of solitons.

\subsection{Interference phenomena}
We find that there are two periodic behaviors in the process of
soliton interaction, one on space distribution and one on time
evolution, which are called by spatial and temporal interference
patterns separately. When the difference between their square velocities or density peaks is large and the relative velocity(RV) is large (how to define large or small will be discussed in next section), the
temporal-spatial pattern will appear, such as Fig.\ 1(a).  When they collide with
different shapes and large RV, the interference pattern will change
in the collision process, such as Fig. 1(b).   Then, one could be
curious about whether there are any definite laws for the nonlinear
interference behavior.

Especially, when solitons collide with identical shapes and square velocities, the temporal pattern will disappear, and just the spatial
interference patterns emerge in Fig. 2.  Keeping solitons' shapes and relative phase unchanged in Fig. 2, we study the
relation between the pattern and RV. It is found that the highest
peak is a constant, and has no relation with RV. However, as the RV
increases,  the spatial period of interference pattern
will decrease. Their
quantitative relation will be discussed in the next section.

\begin{figure}[htb]
\centering
\subfigure[]{\includegraphics[height=65mm,width=80mm]{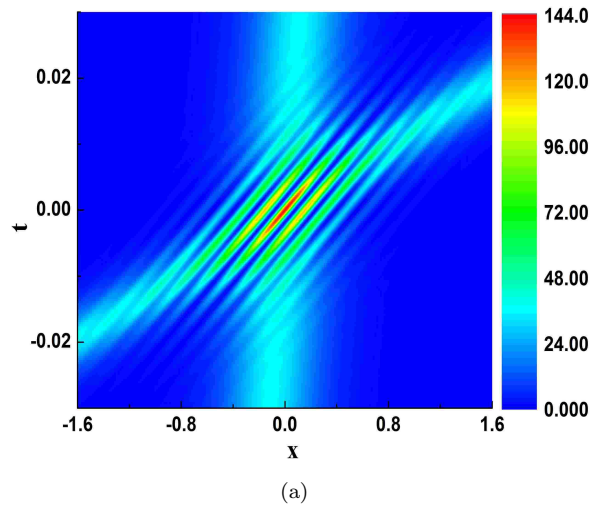}}
\hfil
\subfigure[]{\includegraphics[height=65mm,width=80mm]{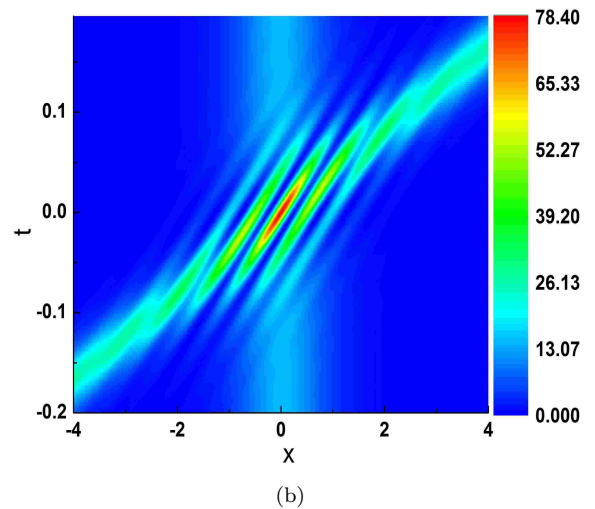}}
\caption{(color online) (a) The temporal-spatial interference
pattern for solitons with same shapes and large kinetic energy
difference. The coefficients are  $a_1=1.5, b_1=1, g=0.25, a_2=1.5,
b_2=20.5$, $c_1=d_1=0$, and $c_2=d_2=0$.  (b) The interference pattern appears
in the collision process of solitons with different shapes and large
RV. The coefficients are $a_1=0.6, b_1=0, g=0.1, a_2=0.8,
b_2=6$, $c_1=d_1=0$, and $c_2=d_2=0$.}
\end{figure}

\begin{figure}[htb]
\centering {\includegraphics[height=65mm,width=80mm]{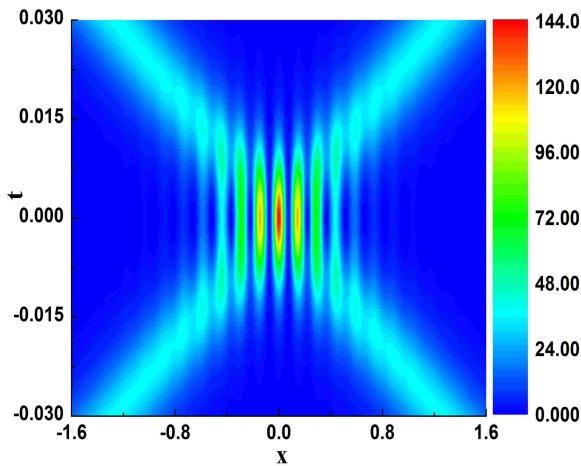}}
\caption{(color online) The density plot of collision region between
the two solitons. It is shown that the spatial interference pattern
appears in the collision region when the relative velocity between
solitons is large. The coefficients are $a_1=1.5, b_1=-10.5, g=0.25,
a_2=1.5, b_2=10.5$, $c_1=d_1=0$, and $c_2=d_2=0$. }
\end{figure}

\subsection{Theoretical analysis}
Believing that some terms related to interference pattern should
appear in the evolution of the two-soliton density, we calculate the
density of the two-soliton solution exactly. The periodic functions
appear in the above solution, which are $\cos(\theta)$ and
$\sin(\theta)$, where
$\theta=2(b_2-b_1)x+4(a_2^2-a_1^2+b_1^2-b_2^2)t$. This means that
there are two periodic behaviors in the evolution of two-soliton,
which refer to spatial and temporal interference patterns
separately. The spatial-period and temporal-period could be
calculated directly as
\begin{equation}
D=\frac{\pi}{b_2-b_1},
\end{equation}
\begin{equation}
T=\frac{\pi}{2(a_2^2-a_1^2+b_1^2-b_2^2)}.
\end{equation}
 But how to understand the physical meaning of the expressions? It is meaningful to derive physical
meaning of each parameter in the two-soliton solution( Eq. (2)), to study the
interference pattern in detail. From the soliton solution, we can
calculate the two solitons' peaks, velocities  through the
asymptotic analysis technic with $b_1\neq b_2$.
 The two solitons' peaks are calculated as
\begin{equation}
P_j=\frac{4 a_j^2}{g},
\end{equation}
and their full width at half maximum $W_j=\frac{1}{2|a_j|}
\ln{(3+2\sqrt{2})}$, which keep unchanged after collision. The
velocity of solitons can be given as
\begin{equation}
v_j= 4 b_j.
\end{equation}
Obviously, $a_j$ and $b_j$ determine
soliton's peak and velocity respectively. One can observe interaction between arbitrary two solitons through varying the parameters. From the asymptotic analysis, as usual, one can know that
the solitons keep their shapes after the collision, and the ``phase shift" emerge \cite{Atre}. Furthermore, we present definite properties of the nonlinear interference period during solitons' interaction process, through combining the density calculation and asymptotic analysis technic.  From Eq. (3), we can know the relative velocity determines the
property of spatial interference pattern, namely
\begin{equation}
D=\frac{4\pi}{v_2-v_1}.
\end{equation}
 From physical viewpoint, soliton could be seen as a particle. In
two-soliton circumstance, one soliton can be chosen as the
reference, the other soliton's velocity will become the RV between
the two solitons. Then, the second soliton's matter wave length will
be determined by RV, based on matter wavelength theory. When RV
increases, the soliton's matter wavelength will decrease. Spatial
interference pattern can be observed when the RV is large enough that
soliton's matter wavelength is smaller than scale of
solitons, such as Fig.\ 1(a). When the matter wavelength is not smaller than the scale of soliton, namely, solitons' relative velocity is small, one could not observe the interference pattern. This is the reason
why spatial interference pattern cannot been seen in the most previous works
 \cite{Akhmediev,Voronvich,Genty,Onorato,Atre}.

From Eq. (4), Eq. (5) and Eq. (6), we can know that the temporal-period is determined by both the peaks and velocities
of solitons, and the nonlinear coefficient, namely,
\begin{equation}
T=\frac{2\pi}{g (P_2^2-P_1^2)+\frac{1}{4}(v_1^2-v_2^2)},
\end{equation}
where $P_j=4a_j^2/g$ ($j=1,2$) denote peaks of solitons.

From the expressions (Eq.\ (7) and Eq.\ (8)) which describe the
periods in space and time, one can manipulate the pattern precisely
through changing the peaks and RV of solitons. When solitons' peaks and square velocities are identical,   the temporal interference pattern will disappear.  Therefore, we can show the
two-soliton just with space-period in Fig. 2. When the velocities of solitons are identical, the spatial interference pattern will disappear. These results provide us particular ways to observe spatial or temporal interference pattern separately.
 When the spatial period less than solitons' scale
and temporal period less than the time scale of collision, we can observe
the temporal-spatial interference pattern in Fig.\ 1. It should be
pointed that the asymptotic analysis based on $b_1\neq b_2$, namely,
the two solitons have different velocities. When two solitons are relatively static, the above analyzes will fail.

\section{Application into nonlinear planar waveguide}
 Experimentally, one can choose spatial optical soliton system to
observe the interference pattern conveniently, due to the
coordinates are both spatial coordinates for spatial optical
soliton. Namely, $t\rightarrow Z$ and $x\rightarrow X$, which are
the transverse and propagation coordinates respectively
\cite{Ku,Neshev}.

\begin{figure}[htb]
\centering {\includegraphics[height=60mm,width=75mm]{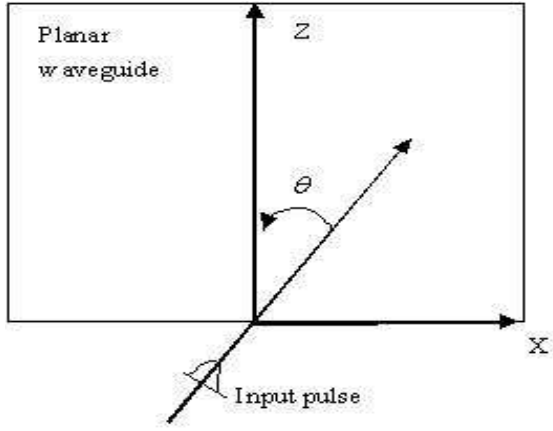}}
\caption{(color online) The incident angle $\theta_j$ is defined as
the rotation angle from the incident direction to $+Z$ direction.}
\end{figure}

 As an example, we consider continuous wave optical
beams propagating inside a planar
 nonlinear waveguide with a refractive index
$$n=n_0+ \gamma I(x,z)\ ,$$
where $I(x,z)$ is the optical intensity, and $x,z$  the transverse
coordinate and propagation distance respectively. Here the first
term $n_0$ stands for the linear part of  refractive index and the
second term represents a Kerr-type nonlinearity. The Kerr
coefficient $\gamma$ can be positive (negative) for nonlinear
self-focusing (self-defocusing) medium. The nonlinear wave equation
governing beam propagation in such a waveguide can be written as
\begin{equation}
i\frac{\partial u}{\partial z}+\frac{1}{2k_0}\frac{\partial^2
u}{\partial x^2}+\frac{k_0\gamma}{n_0}|u|^2u=0,
\end{equation}
where $k_0=2\pi n_0/\lambda_0$ is the wave number of the optical
source generating the beam. After introducing normalized variables
$U=u,~X=\sqrt{2k_0}x,$ $Z=z$, and $2g=\frac{k_0\gamma}{n_0}$, the
Eq.(8) can be rewritten as
\begin{equation}
i\frac{\partial U}{\partial Z}+\frac{\partial^2 U}{\partial X^2}+ 2g
|U|^2U=0
\end{equation}
In this system, the velocities of solitons will denote the tangent values of incident angles. The incident angle is defined as
the rotation angle from the incident direction to $+Z$ direction, as shown in Fig. 3.
The RV before will become the
difference between the tangent values of incident angles. Therefore, the
interference pattern could be observed much more conveniently in the
nonlinear planar waveguide. For this system, the parameter $g$ is
related to Kerr nonlinear parameter. We show that the nonlinear
parameter can be derived from the interference pattern with some
initial conditions of solitons. As an example, we study interference
pattern in Fig. 1(b). The corresponding initial solitons are shown
in Fig. 4(a). Their intensity values $P_j$ are assumed to be known,
which can be measured more easily. The incident angle of one soliton
is known to be zero, namely $\theta_1=0$. From the interference
pattern, one can measure the periods along the two directions
independently, as shown in Fig. 4(b) and (c). Then the other
soliton's incident angle can be known as
\begin{equation}
\theta_2=arctan [\frac{4\pi}{D}].
\end{equation}
Furthermore, the parameter $g$ will be given by
\begin{equation}
g=\frac{2\pi}{ T (P_2^2-P_1^2)}+\frac{4 \pi^2}{D^2(P_2^2-P_1^2)}.
\end{equation}

\begin{figure}[htb] \centering
\subfigure[]{\includegraphics[height=50mm,width=70mm]{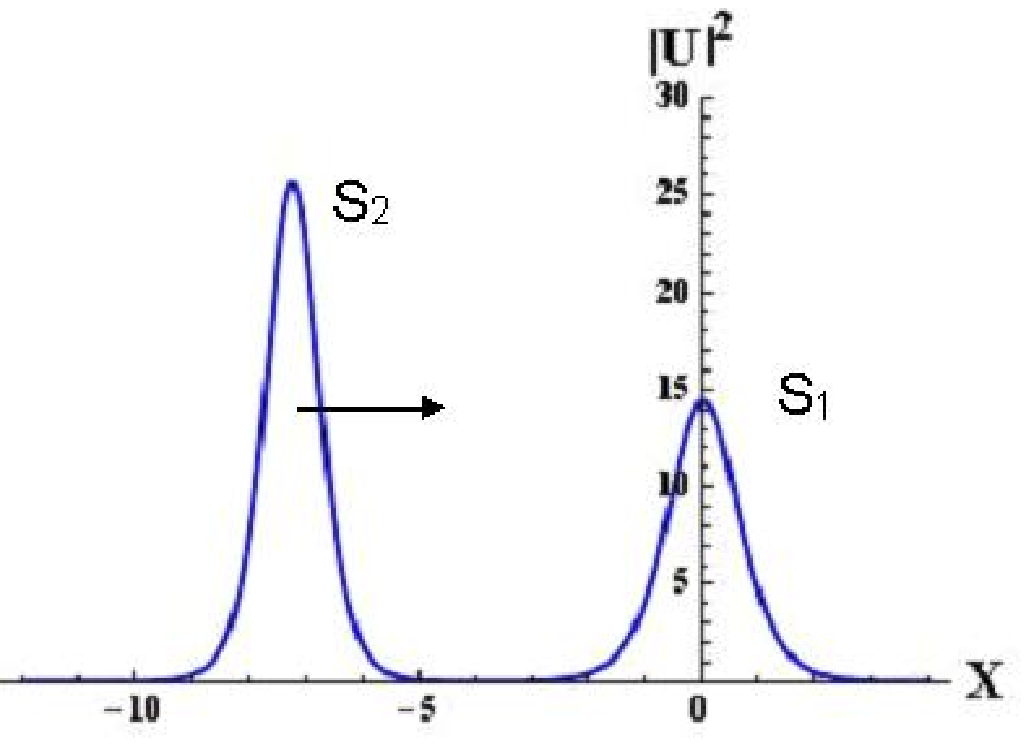}}
\hfil
\subfigure[]{\includegraphics[height=30mm,width=40mm]{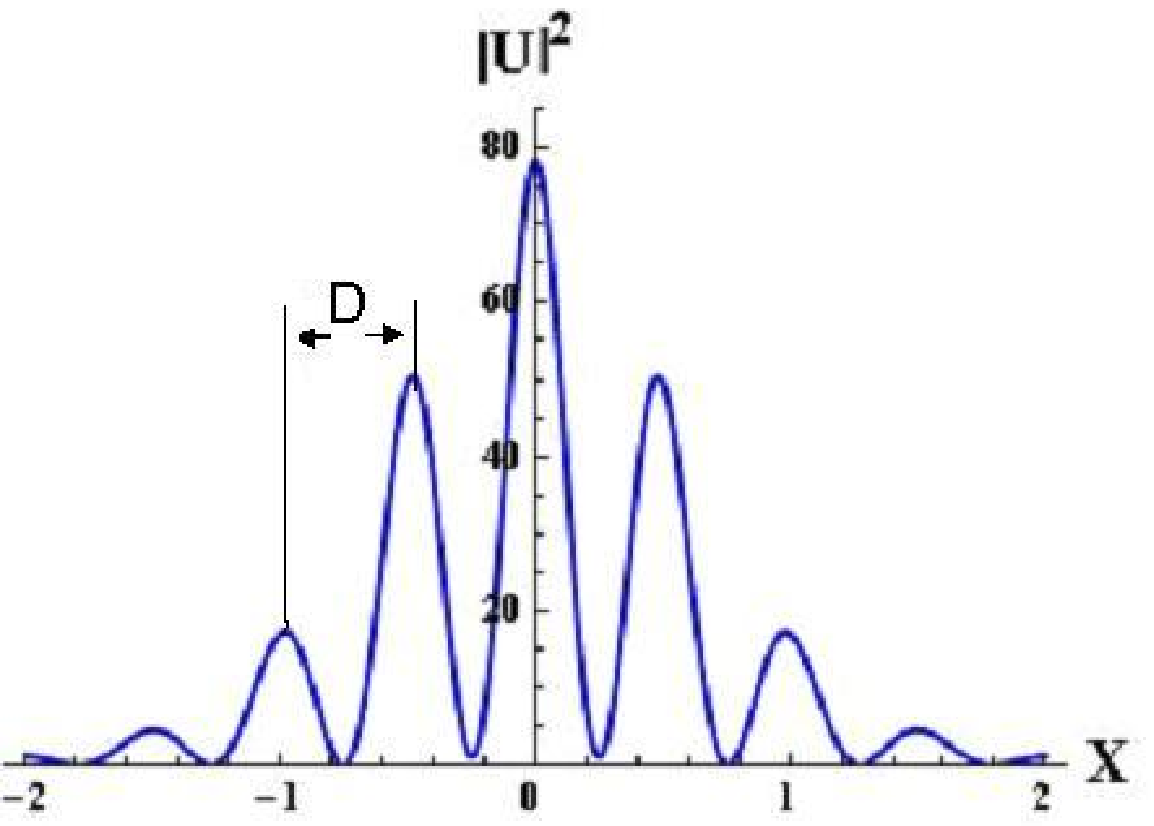}}
\hfil
\subfigure[]{\includegraphics[height=30mm,width=40mm]{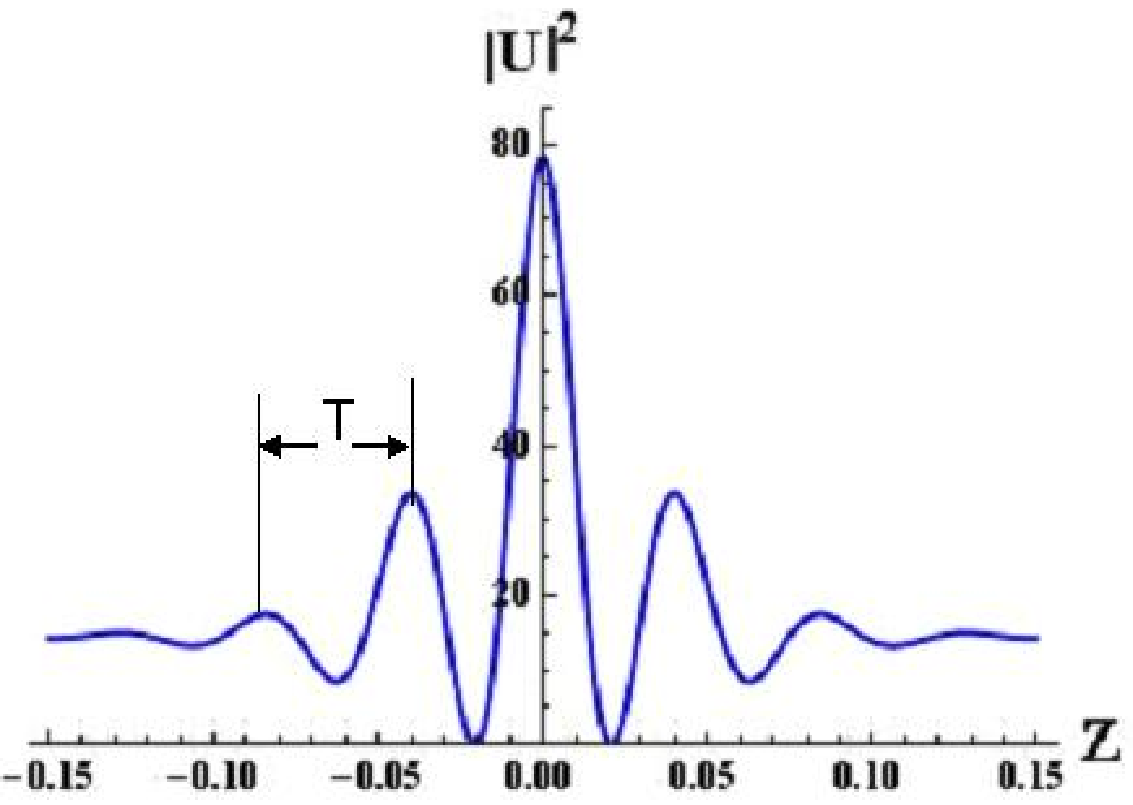}}
\caption{(color online) (a) The initial shape of the two solitons in Fig. 1(b)
before collision.
(b) The period $D$ on transverse direction. (c) The period $T$ on propagation direction.  They are the cutaway
view of the interference pattern in Fig. 1(b) at $Z=0$ and $X=0$
respectively. }
\end{figure}

 Based on the relation, we
can calculate $g$ from the period of longitudinal interference
pattern and initial information of solitons before collision. Then,
the Kerr nonlinear coefficient can be obtained easily from
$\gamma=\frac{2g n_0 }{k_0}$, for $k_0$ can be known from the wave
number of the optical source generating the beam, $n_0$ is very easy
to be measured for it is linear refractive index. Similar discussion
can be made on solitons in BEC, for the dynamic equations are
identical fundamentally \cite{Zhao,B. Wu,G.X. Huang,M. Matuszewski}. Namely, we find a possible way to measure
nonlinear parameter through observing the interference pattern. It
should be noted that the expressions will fail to predict the
nonlinear parameter when the peak values of solitons are identical.
However, the width and peak of solitons should be related in a
certain way, $g= \frac{[\ln(3+2\sqrt{2})]^2}{P_j W_j^2}$. Namely,
one can calculate the nonlinear coefficient from the information of
soliton's shape too.

\section{Conclusion}
 It should be emphasized that the collision between solitons is still elastic. Here, we
 focus on the process of solitons' collision. We find that temporal-spatial interference pattern can
appear during the interaction of solitons. Period of spatial
interference pattern is determined by the relative velocity of
solitons. Spatial interference fringes can be
just observed under the condition that soliton's matter wavelength
is shorter than the scale of solitons. Temporal interference pattern
can be seen when two solitons are relatively static and the distance
between them is not large. From the expressions which describe the
period of the pattern, we know that interference pattern could be
manipulated precisely through changing RV and intensity peaks of solitons. As
an example, we show the possibilities to observe the patterns in a
nonlinear planar waveguide. We find that the Kerr nonlinear
coefficient can be read out from the interference pattern, with two
solitons' density peaks and incident angles known.

\section*{Acknowledgments}
 We are grateful to Professor C.T. Zhou for helping in theoretical analysis. This work
is supported by the National Fundamental Research Program of China
(Contact No. 2011CB921503), the National Science Foundation of China
(Contact Nos. 11075020, 11078001).

\end{document}